\documentclass{article}

\PassOptionsToPackage{square,sort,comma,numbers}{natbib}

\usepackage[final]{neurips_2019}

\usepackage[utf8]{inputenc} 
\usepackage[T1]{fontenc}    
\usepackage{hyperref}       
\usepackage{url}            
\usepackage{booktabs}       
\usepackage{amsfonts}       
\usepackage{nicefrac}       
\usepackage{microtype}      
\setcitestyle{square}
\usepackage{epsfig}
\usepackage{graphicx}
\usepackage{floatrow}
\usepackage{subcaption}
\usepackage{sidecap}

\title{Multi-defect microscopy image restoration under limited data conditions}

\author{%
  Anastasia Razdaibiedina\\
  Vector Institute\\
  University of Toronto\\
  \texttt{anastasia.razdaibiedina@mail.utoronto.ca} \\
   \And
   Jeevaa Velayutham \\
   University of Toronto \\
   \texttt{jevaav@cs.toronto.edu} \\
   \AND
   Miti Modi \\
   University of Toronto \\
   \texttt{miti.modi@mail.utoronto.ca} \\
}

\begin{document}

\maketitle

\begin{abstract}
  Deep learning methods are becoming widely used for restoration of defects associated with fluorescence microscopy imaging. One of the major challenges in application of such methods is the availability of training data. In this work, we propose a unified method for reconstruction of multi-defect fluorescence microscopy images when training data is limited. Our approach consists of two steps: first, we perform data augmentation using a Generative Adversarial Network (GAN) with conditional instance normalization (CIN); second, we train a conditional GAN (cGAN) on paired ground-truth and defected images to perform the restoration. The experiments on three common types of imaging defects with different amounts of training data, show that the proposed method gives comparable results or outperforms CARE, deblurGAN and CycleGAN in restored image quality when limited data is available.
\end{abstract}

\section{Introduction}
Degradation of fluorescence microscopy images occurs due to many factors, including bleaching of fluorophores, readout noise from sensors, out-of-focus light, poor axial sampling, and uneven illumination and detection \cite{belthangady2019applications}. Microscopy hardware constraints and physical limitations often make it impossible to simultaneously overcome multiple types of imaging defects. Hence, the application of image reconstruction pipelines is becoming increasingly popular. In this work we focus on three common tasks in microscopy image restoration: \textit{denoising}, \textit{axial inpainting}, and deep learning enabled \textit{super-resolution} \cite{luisier2010image, belthangady2019applications, schermelleh2019super}.

We used three publicly available datasets, which represent the above-mentioned defect types (Table \ref{data}) \cite{Weigert2018ContentawareIR}. For image quality assessment, we used Peak Signal-to-Noise Ratio (PSNR) and Structural Similarity index (SSIM) scores \cite{Wang2004ImageQA}.

\begin{table}[htb]
\begin{center}
\begin{tabular}{|c|c|c|c|c|}
\hline
Organism & Biological Structure & Restoration task & average PSNR / SSIM\\
\hline
Drosophila & whole embryo & denoising & 19.92 / 0.63 \\
Zebrafish & retinal nuclei & axial inpainting & 12.51 / 0.33 \\
Human & HeLa cells microtubules & super-resolving & 7.27 / 0.08 \\
\hline
\end{tabular}
\end{center}
\caption{Description of datasets used for the experiments. PSNR / SSIM column denotes average PSNR and SSIM scores across the whole dataset before image reconstruction.}
\label{data}
\end{table}

Considering the diversity of reconstruction tasks in fluorescent microscopy, and due to an increasing need in data-driven image restoration approaches, we propose a unified method for multi-defect micrograph restoration. To overcome the problem of limited training data, which is a common challenge for experimental databases, we incorporate meta-learning style data augmentation into our pipeline.

\section{Method}

\begin{figure}

a)
  \includegraphics[width=0.95\textwidth]{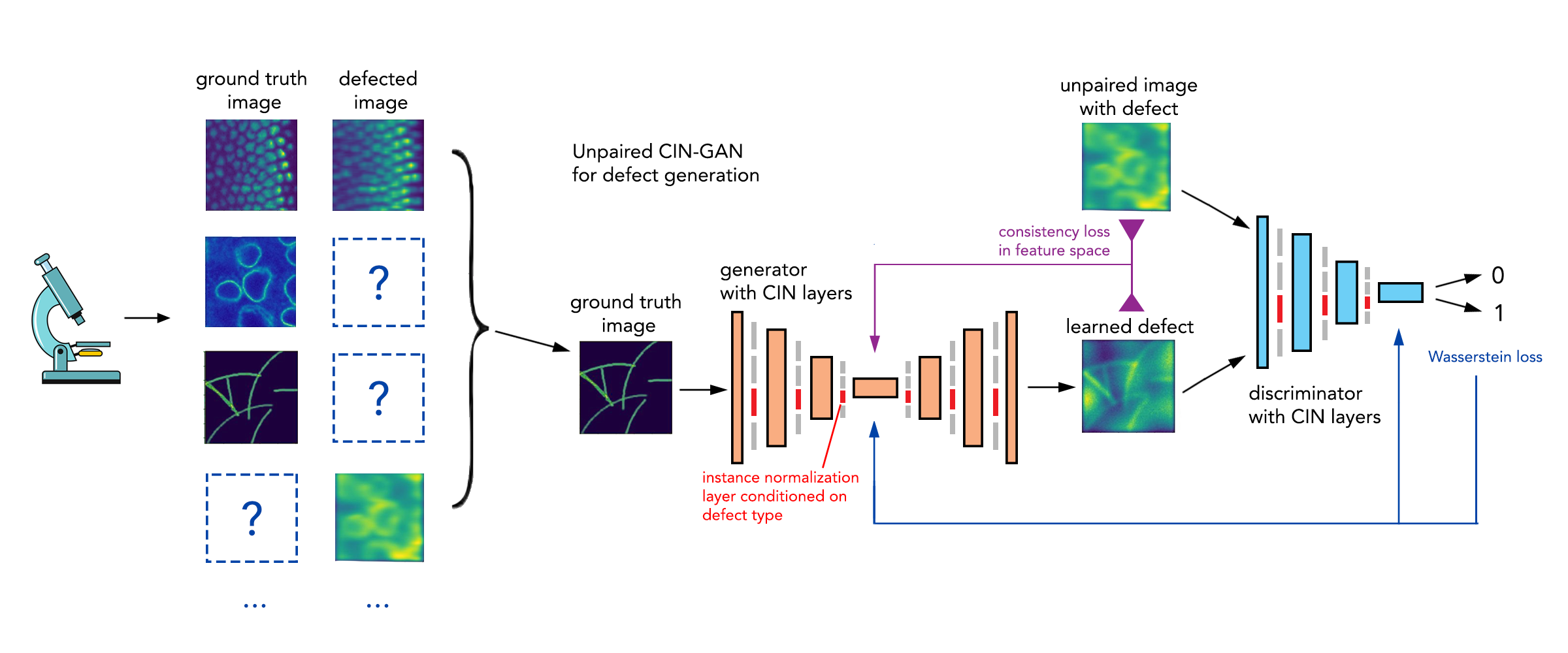}

b)
  \includegraphics[width=0.95\textwidth]{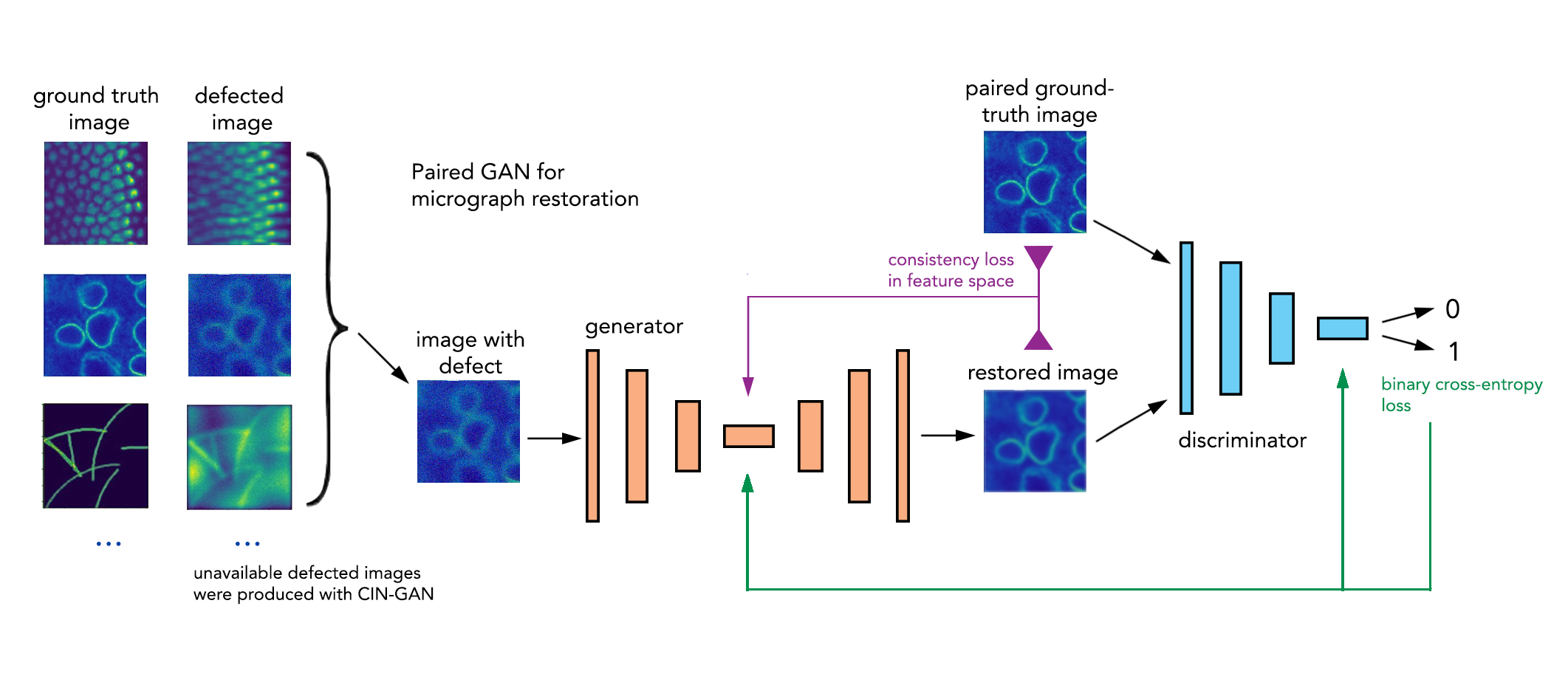}

\caption{Illustration of the proposed pipeline. a) GAN with conditional instance normalization (CIN) layers for generating different microscopy defects. Since paired data is often unavailable, CIN-GAN is trained on a small amount of unpaired defected and ground truth microscopy image data; different CIN layers are "switched" on / off depending on the task. b) Conditional GAN for image restoration. This GAN is trained on paired data, where the previously unavailable defected image pairs are now produced by CIN-GAN.}

\end{figure}

The proposed pipeline consists of two GANs \cite{Goodfellow2014GenerativeAN}. First, a GAN with CIN layers is trained to perform data augmentation using limited amount of unpaired ground-truth and defected images. We apply condition on the instance normalization for each defect type as proposed by \cite{Dumoulin2017ALR}:
\begin{equation} z = \gamma_{i}(\frac{x-\mu}{\sigma}) + \beta_{i}
\end{equation}
where $i$ is the number of tasks, $\mu$ and $\sigma$ are mean and standard deviation respectively of the input. The parameters $\gamma_{i}$ and $\beta_{i}$ are learned separately for each task. Thus, all other layers in the model are shared across tasks, except for the instance normalization layers that drive distinct defects synthesis. Second, the dataset is augmented by CIN-GAN and a cGAN \cite{isola2017image} is trained on paired high-resolution ground-truth images and defective images. The resulting cGAN is used to restore multiple types of microscopy defects. For training of both GANs we used adversarial and content losses, where content loss measures image consistency in feature space of VGG16 model \cite{Gatys2015ANA, simonyan2014very}:
\begin{equation}
L_{total} = L_{adv} + \lambda \cdot L_{content}
\end{equation}

\section{Results}

\begin{figure}
     \centering
     \begin{subfigure}[b]{0.32\textwidth}
         \centering
         \includegraphics[width=\textwidth]{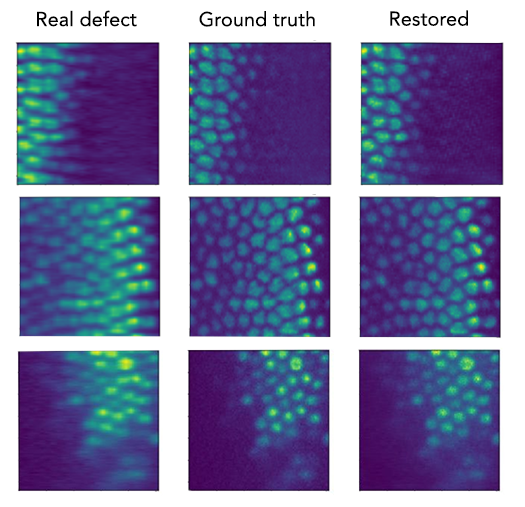}
         \caption{Denoising of Drosophila embryonic cells.}
         \label{fig:y equals x}
     \end{subfigure}
     \hfill
     \begin{subfigure}[b]{0.32\textwidth}
         \centering
         \includegraphics[width=\textwidth]{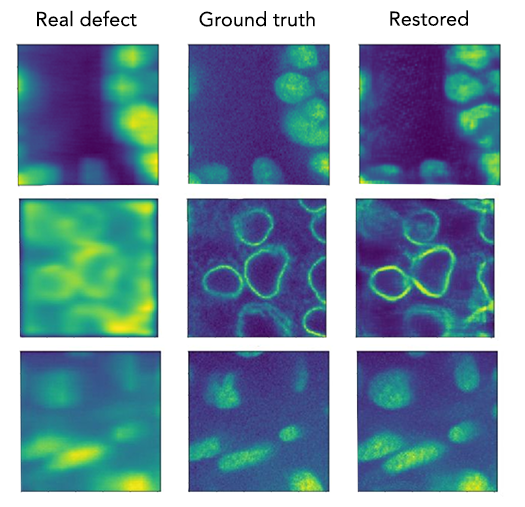}
         \caption{Axial inpainting of Zebrafish retinal nuclei.}
         \label{fig:three sin x}
     \end{subfigure}
     \hfill
     \begin{subfigure}[b]{0.32\textwidth}
         \centering
         \includegraphics[width=\textwidth]{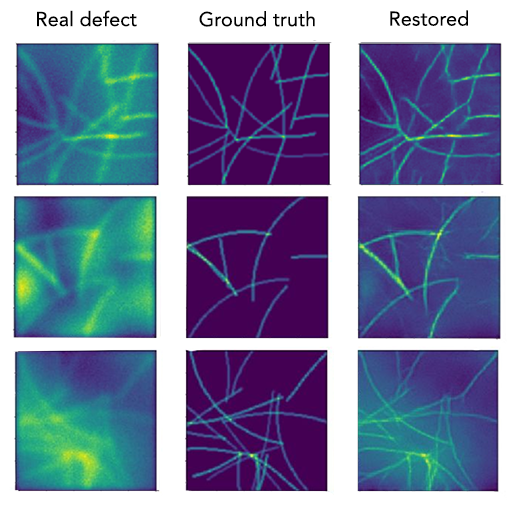}
         \caption{Super-resolving microtubules of HeLa cells.}
         \label{fig:five over x}
     \end{subfigure}
        \caption{Examples of micrographs restoration with the proposed pipeline on three distinct tasks. 10 real pairs of images were used for training; dataset was augmented with CIN-GAN.}
        \label{images}
\end{figure}

We compared performance of our restoration pipeline in two settings: A) a limited amount of real paired images (defected + in-focus) are available; B) no real pairs are available, all defected pairs for cGAN training were synthetically generated. Restoration results under setting of 10 paired images are shown in Figure \ref{images}.

A) For the paired setting, we compared our model performance with two existing state-of-the-art image restoration models: (1) DeblurGAN \cite{Kupyn2018DeblurGANBM}, (2) U-Net based \cite{ronneberger2015u} CARE models \cite{Weigert2018ContentawareIR}. These two models are trained on paired images, that is, we assume that there are real paired images available for model training. We outperformed both models for most tasks from 10 to 50 pairs of images, results for 10 pairs of images are shown in Table \ref{paired} and Figure \ref{images}. B) For the unpaired setting we assumed that real paired training data was not available. CIN-GAN was used to generate artificial defects to create paired training data for cGAN which restores the real defected images. We compare this performance with state-of-the-art unpaired image to image translation model, CycleGAN \cite{Zhu2017UnpairedIT} in Table \ref{paired}.

\begin{SCfigure}
  \centering
  \includegraphics[width=60mm]{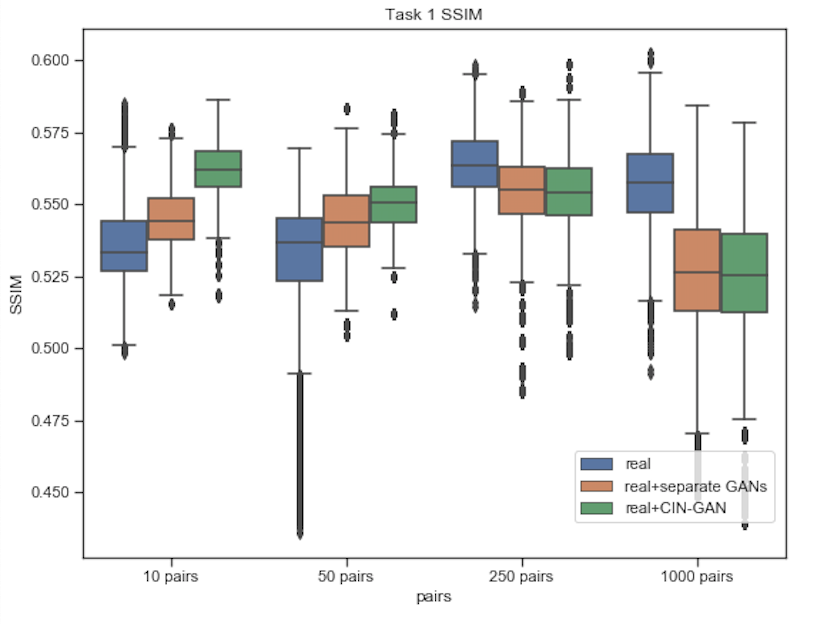}

   \caption{Benefit of GAN-based data augmentation. Boxplots show the distribution of SSIM scores for restored images for real paired data, real data augmented by separate GANs and real data augmented by CIN-GAN. Plots are shown for 10, 50, 250 and 1000 data pairs. CIN-GAN data augmentation proves to be beneficial in cases when limited paired data is available.}
\label{fig:main_plot}
\end{SCfigure}

We have also shown advantage of using CIN-GAN in the limited data setting by comparing performance of data augmentation via CIN-GAN and separately trained GANs (Figure \ref{fig:main_plot}).

\begin{table}[htb] 
\begin{center}
\begin{tabular}{|c|c|c|c|}
\hline
& Denoising & Axial inpainting & Super-resolution \\
\hline\hline
CARE & 21.6/\textbf{0.56} & 12.8/0.29 & 14.1/0.20 \\
DeblurGAN & 18.0/0.33 & 14.6/0.20 & 11.2/0.14 \\
Ours & \textbf{22.4}/\textbf{0.56} & \textbf{17.4}/\textbf{0.38} & \textbf{14.3}/\textbf{0.22} \\
\hline
CycleGAN & 21.3/0.49 & \textbf{15.3}/0.27 & \textbf{12.7}/\textbf{0.14} \\
Ours & \textbf{21.9}/\textbf{0.57} & 15.1/\textbf{0.32} & 8.9/0.07 \\
\hline
\end{tabular}
\end{center}
\caption{Comparison of PSNR/SSIM scores between our cGAN restoration network with other models with 10 paired images (top) or no paired data (bottom).}
\label{paired}
\end{table}

\sidecaptionvpos{figure}{c}

Thus, we proposed a unified method for microscopy image restoration that is applicable for reconstruction of multiple defect types under limited data conditions. With the help of GAN-based data augmentation we are able to overcome the problem of unpaired limited data. The proposed pipeline achieves competitive results on three microscopy databases and can be scaled up to more defect types.

\bibliographystyle{ieeetr}
\bibliography{egbib}

\end{document}